\documentclass[pre,twocolumn,showpacs,aps,superscriptaddress]{revtex4}
\usepackage{graphicx,epsfig}
\usepackage{rotate}
\usepackage{color}
\usepackage{amsmath,amssymb}

\begin{document}

%\preprint{APS/123-QED}

\title{Analytical estimation of the correlation dimension of integer lattices}% Force line breaks with \\

\author{Lucas Lacasa}
\email{l.lacasa@qmul.ac.uk}
\affiliation{School of Mathematical Sciences, Queen Mary University of London,
Mile End Road, E14NS London, UK}%

\author{Jes\'us G\'omez-Garde\~nes}
\email{gardenes@gmail.com}
\affiliation{Institute for Biocomputation and Physics of Complex System (BIFI)}
\affiliation{Departamento de Fisica de la Materia Condensada, Universidad de Zaragoza, Spain}%

\date{\today}% It is always \today, today,

\begin{abstract}
Recently [L. Lacasa and J. G\'omez-Garde\~nes, Phys. Rev. Lett. {\bf 110}, 168703 (2013)], a fractal dimension has been proposed to characterize the geometric structure of networks. This measure is an extension to graphs of the so called {\em correlation dimension}, originally proposed by Grassberger and Procaccia to describe the geometry of strange attractors in dissipative chaotic systems. The calculation of the correlation dimension of a graph is based on the local information retrieved from a random walker navigating the network. In this contribution we study such quantity for some limiting synthetic spatial networks and obtain analytical results on agreement with the previously reported numerics. In particular, we show that up to first order the correlation dimension $\beta$ of integer lattices $\mathbb{Z}^d$ coincides with the Haussdorf dimension of their coarsely-equivalent Euclidean spaces, $\beta=d$.  
\end{abstract}

%\pacs{}% PACS, the Physics and Astronomy
\maketitle

\noindent {\bf In this article we address the concept of correlation dimension which has been recently extended to network theory in order to efficiently characterize and estimate the dimensionality and geometry of complex networks \cite{PRL}. This extension is inspired in the Grassberger-Procaccia method \cite{seminal1,seminal2,seminal3}, originally designed to quantify the fractal dimension of strange attractors in dissipative chaotic dynamical systems. When applied to networks, it proceeds by capturing the trajectory of a random walker diffusing over a network with well defined dimensionality. From this trajectory, an estimation of the network correlation dimension is retrieved by looking at the scaling of the walker's correlation integral. Here we give analytical support to this methodology by obtaining the correlation dimension of synthetic networks representing well-defined limits of real networks. In particular, we explore fully connected networks and integer lattices, these latter being coarsely-equivalent  \cite{coarse} to Euclidean spaces. We show that their correlation dimension coincides with the the Haussdorff dimension of the respective coarsely-equivalent Euclidean space.}
\section{Introduction}

During the last decade the science of networks has shed light on the importance that the real architecture of the interactions among the constituents of complex systems has on the onset of collective behavior \cite{rev:albert,rev:newman,rev:bocc}. In this way it has contributed to the advance in many branches of science, such as statistical physics and nonlinear dynamics, in which the understanding of collective phenomena is fundamental. While the structural aspects of networks have been largely explored by means of topological measures \cite{newmanbook}, their geometrical aspects have been ignored, with the remarkable exception of a few attempts to characterize the dimensionality of their complex interaction backbone \cite{emilio,boxcovering, serrano,serrano1}. For instance, the box-counting technique, widely used for estimating the capacity dimension $D_0$ of an object, was extended in \cite{boxcovering, boxcovering2, boxcovering3, jstat} as a box-covering algorithm, aimed at characterizing the dimensionality of complex networks. 

Recently \cite{PRL}, we proposed an extension of the concept of correlation dimension \cite{review} to estimate the dimensionality of complex networks by using random walkers to explore the network topology. This extension builds up on the well-known Grassberger-Procaccia method \cite{seminal1,seminal2,seminal3}, originally designed to quantify the fractal dimension of strange attractors in dissipative chaotic dynamical systems. This approach relies on embedding a trajectory of the dynamical system in an \textit{m}-dimensional space and calculating a correlation integral over this trajectory. 

The rationale of the extension of the Grassberger-Procaccia method to the network realm is that the geometrical structure of the network restricts the movement of a random walker and, accordingly, a notion of dimensionality can be extracted through the properties of the walker's trajectory. In particular, if the trajectory evolves over some object with well-defined correlation dimension, such dimension, $\beta$, should be accessible experimentally through the scaling of the walker's correlation sum defined in the next section. 
In addition to its novelty, the use of the Grassberger-Procaccia method together with the machinery of random walks, provides another nice example of the use of walkers to capture the structure and organization of a complex network, such as the centrality of nodes \cite{pr}, its community structure \cite{rosvall} or the existence of degree correlations \cite{vito}.

In \cite{PRL} we showed numerical estimates of the correlation sum for walkers navigating a set of synthetic and real-world networks, finding a range of dimensions $1<\beta<3$ (comprising integer and fractal values) for systems such as the world-wide air transportation network, road and energy networks or the Internet. On the other hand, other networks lack a scaling for the correlation sum, distinguishing those systems whose structure has a strong degree of self-similarity from others in which such fundamental symmetry is missing. 
In the present contribution we give some analytical support to the findings and conjectures shown in \cite{PRL}. We first address fully connected networks, which intuitively can only be embedded in infinite dimensional spaces, and show that the correlation dimension is indeed a diverging quantity. Then we address integer lattices, which are coarsely-equivalent \cite{coarse} to Euclidean spaces, giving analytical evidence that their correlation dimension coincides with the Haussdorff dimension of the respective coarsely-equivalent Euclidean space.

\section{Correlation Dimension from Random Walks in Networks}

Let us start  by briefly reviewing the generalization of the Grassberger-Procaccia method to the computation of the correlation dimension of a complex network.  We denote by  ${\cal G}$ a spatially embedded undirected network with $N$ nodes and $L$ links, so that each node $i$ of ${\cal G}$ is labelled by a generic vector $\textbf{v}_i$ that uniquely determines the location of node $i$ in the underlying space ($\textbf{v} \in \mathbb{R}^d, \  \text{or} \in \mathbb{Z}^d$ when the space is discrete). The network topology is given by the so-called $N\times N$ adjacency matrix ${\bf A}$, whose elements are defined (for undirected and unweighted graphs) as $A_{ij}=A_{ji}=1$ when nodes $i$ and $j$ are connected and $A_{ij}=A_{ji}=0$ otherwise.

Once the network is defined, we must define the dynamical evolution of a random walker on network ${\cal G}$. The time-discrete version of a random walks determines that, at each time step $t$, the walker at some node $i$ hops to one of the neighbors $j$ with equal probability. In this way the transition matrix ${\bf M}$ of a walker defines the probability that a walker at node $i$ at time $t$ is at a node $j$ at time $t+1$ as:
\begin{equation}
M_{ij}=\frac{A_{ij}}{\sum_{l=1}^N A_{il}}=\frac{A_{ij}}{k_i}\;,
\label{markov}
\end{equation}
where $k_i=\sum_{l=1}^N A_{il}$ is the degree of node $i$. Thus, by initially setting the a walker at some randomly chosen node,  one iterates the dynamics prescribed by matrix ${\bf M}$ and follows the trajectory of the walker (note at this point that in practice one does need to store ${\bf M}$, as we only need to have (local) information at each time step of the neighbors of a given node, rendering this method useful for practical situations involving arbitrarily large networks, e.g. Internet).

Now consider a trajectory of length $n$ generated by an ergodic random walker navigating the network ${\cal G}$ as described above. The trajectory can be described as the sequence of visited nodes. In the case of spatial networks, the trajectory can be casted in the series $\{{\bf v}_1,{\bf v}_2,\dots,{\bf v}_n\}$, and embed the series in $\mathbb{R}^{m\cdot d}$ (where $m$ is the embedding dimension) by defining the vector-valued series $\{{\bf V}(t)\}$, where ${\bf V}(i) \in \mathbb{R}^{m\cdot d}$ is defined as:
\begin{equation}
{\bf V}(i)=[\textbf{v}_{i+1},\dots,\textbf{v}_{i+m-1}].
\end{equation}
%(note that the delay time in the original embedding method is set here to $\tau=1$ since in the network context large jumps are allowed).

Finally, the correlation sum function $C_m(r)$ is defined as the fraction of pairs of vectors whose distance is smaller than some similarity scalar $r \in \mathbb{R}$:
\begin{equation}
C_m(r)=\frac{2\sum_{i<j} {\Theta}(\|{\bf V}(i) - {\bf V}(j)\|-r)}{(n-m)(n-m+1)}\;,
\label{CS}
\end{equation}
where $\Theta(x)$ is the Heaviside step function, and $\|\cdot\|$ is a p-norm $\|{\bf x}\|_p= \bigg[\sum_i |x_i|^p\bigg]^{1/p}$. Here, without loss of generality, we choose for convenience $\|\cdot\|$ as the $L_\infty$ norm, $\|{\bf x}\|_\infty= \max( |x_1|,|x_2|,\dots,|x_n| )$. Note, that  although within the seminal Grassberger-Procaccia method the use of the Euclidean norm was proposed \cite{seminal1,seminal2,seminal3}, the use of $L_{\infty}$ norm was later adopted by Takens in \cite{takens1}, although the results obtained should be norm invariant \cite{review}.

The main scaling conjecture that was proposed and addressed numerically in \cite{PRL} states that when the series is extracted from the trajectory of a random walker navigating a network ${\cal G}$ with \textit{well defined dimension}, for sufficiently long series and sufficiently small values of $r$, $C_m(r)$ evidences a scaling regime such that:
\begin{equation}
\lim_{r\rightarrow0}\lim_{n\rightarrow\infty}\frac{\log(C_m(r))}{\log(r)}=\beta_m\;.
\label{beta}
\end{equation}
The value $\beta_m$ approaches a constant value $\beta_m\rightarrow \beta$ for sufficiently large embedding dimension $m$. This latter value $\beta$ constitutes the estimate of the network's 'correlation dimension'. Notice that, in practice, the limit $r\rightarrow 0$ should be substituted by a sufficiently small $r$ which depends on the characteristic space labeling, {\em i.e.}, if nodes are labelled by integer valued vectors then the limit $r\rightarrow 0$ should be substituted by $r\ll n$.

\noindent 
\section{Fully connected network}

After introducing the basis for the calculation of the correlation dimension of spatial graphs we begin our study with the simple case of a fully connected network. This network, also termed as complete, is a graph {\cal G} in which each node is connected with the rest of the $N-1$ nodes and thus the adjacency matrix reads $A_{ij}=1-\delta_{ij}$ (with $\delta_{ij}=1$ if $i\neq j$ and $\delta_{ii}=0$). Note that the fully connected network can be understood as the dense-limit ($L\rightarrow N^2$) of a real network.

A fully connected network can be embedded in an Euclidean space with diverging dimensionality, where each node $i$ is in turn labeled  by an infinite dimensional vector:
 \begin{equation}
 \textbf{X}_i=\sum_{i=1}^\infty \alpha_i \textbf{e}_i\;,
 \end{equation} 
with $\alpha_i \in \mathbb{N}$ and $\textbf{e}_i \cdot \textbf{e}_j=\delta_{ij}$. In order to prove that the correlation dimension $\beta$ of such object diverges, we need to find that $\beta_m$ is a monotonically increasing function of the embedding dimension $m$. 
 
In what follows we prove the above claim. First, notice that the transition matrix \textbf{M} (Eq. (\ref{markov})) of a random walker navigating a fully connected network reads: 
 \begin{equation}
 \text{M}_{ij}=\frac{1-\delta_{ij}}{N-1}\;.
  \end{equation} 
Showing that the walker can hop between any pair of nodes $i$ and $j$ with equal probability. This makes the infinite dimensional labeling above arbitrary for any practical purpose. Thus for convenience and without loss of generality, we label each node by a random number $x$ extracted from a uniform distribution $U[0,1]$. Accordingly, a random walker navigating this fully connected network generates a trajectory which is a sequence of $n$ independent and identically distributed random variables, $\{x(1),x(2),x(3),\dots,x(n)\}$, where each $x(i)\in U[0,1]$.

Consider now the embedding vector as a positive-definite random variable itself, {\em i.e.},
 \begin{equation}
 \Vert \textbf{V}(i)-\textbf{V}(j) \Vert \equiv \xi\;,
\end{equation} 
extracted from some unknown probability density $\xi \in \rho(x), x\geq0$. After dropping irrelevant constants, the correlation sum (Eq. \ref{CS}) reduces to the probability:
 \begin{equation}
P(\xi < r)=\int_0^r \rho(x)dx\;.
  \end{equation} 
Our program is based on the calculation of $\rho(x)$.\\

Let us begin with embedding dimension $m=1$. In this case $\textbf{V}(i)=\textbf{v}_i=x(i)$ and, according to the $L_{\infty}$ norm: 
 \begin{equation}
\xi=|x(i)-x(j)|\;,
  \end{equation} 
where we recall that $x(i)$ and $x(j)$ are uniformly distributed random variables. Trivially, $\xi$ is distributed according to a triangular distribution $f(x)=2(1-x)$. Hence $\rho(x)=f(x)$ and
 \begin{equation}
C_1(r)\sim P(\xi<r)=\int_0^r (2-2x)dx=2r-r^2\;.
  \end{equation} 
For small values of $r$, the scaling is linear, and we obtain:
 \begin{equation}
\lim_{r \rightarrow 0}  \frac{\log[C_m(r)]}{\log r}= 1 + \text{h.o.t.}\;,
 \end{equation} 
that is, up to first order we find $\beta_1=1$.

In a second step let us consider the case $m \geq 2$, for which $\textbf{V}(i)=(x(i),x(i+1),\dots, x(i+m-1))$, for which
 \begin{equation}
 \xi= \max \{|x(i+l)-x(i+l+\alpha)|; l=0,...,m-1 \}\;, 
 \end{equation} 
where each of the random variables of the form $|x(i)-x(j)|$ is now distributed following a triangular distribution $f(x)$.
Our problem thus lies in deriving how $\xi$ is distributed. Note that this problem reduces to an extreme value problem, which can be solved using order statistics such that:
 \begin{equation}
 \xi \sim m f(x)[F(x)]^{m-1}\equiv \rho(x)\;,
 \end{equation} 
where $F(x)=\int_0^x f(x)dx$ is the cumulative distribution function of $f(x)$. Therefore, in this general case the correlation sum yields:
 \begin{eqnarray}
 C_m(r)\sim P(\xi<r)&=&\int_0^r 2m(1-x)(2x-x^2)^{m-1}dx\nonumber
 \\
 &\sim& r^m + \text{h.o.t.}\;,
 \end{eqnarray}
Thus, we conclude that, up to first order, the correlation sum of a random walker navigating a fully connected network evidences a so called trivial scaling with the similarity distance $r$: the exponent of the scaling  $\beta_m$   increases linearly with the embedding dimension without saturation, $\beta_m = m$. This result is reminiscent of the infinite dimensional attractor of white noise in the original Grassberger-Procaccia procedure \cite{seminal1,seminal2,seminal3}, and, applied to the network realm, it corresponds to an infinite correlation dimension. $\square$\\

\section{Integer lattices}

In what follows we address integer lattices $\mathbb{Z}^d$, which are coarsely-equivalent \cite{coarse} to Euclidean spaces with Haussdorff dimension $d$. For $d\leq2$ these lattices are, for instance, the regular-limit of road or infrastructure networks (in this limit, all nodes have the same degree $k_i=2^d \ \forall i=1,...,N$ and are homogeneously located in the underlying space, tiling it in a regular way), and for $d\geq 3$ these lattices respresent discretizations of the Euclidean space.

\subsection{1D Lattice}
A 1D lattice is simply a chain graph which, intuitively, tends to an object of Haussdorff dimension one as the distance between nodes shrinks continuously to zero. In what follows we propose two alternative proofs, a ballistic approximation and a calculation based on the unbiased motion of random walkers, both showing that the correlation dimension of 1D lattices is $\beta=1$.

\subsubsection{Ballistic approximation}

As an approximation (relaxed below), let us first consider the case of a ballistic (deterministic) walker in the 1D lattice. If this lattice is labeled without loss of generality by integers (where two adjacent nodes are labeled as $i$ and $i+1$, and $A_{ij}=\delta_{i,i+1}+\delta_{i,i-1}$), then a typical walker produces the string $\{i,i+1,i+2,i+3,i+4,...\}$ or, by symmetry $\{i,i-1,i-2,i-3,i-4,...\}$.
Both cases are equivalent and therefore yield equivalent results. We shall therefore address the former for concreteness.

Let us start with embedding dimension $m=1$. Then,
\begin{equation}
\Vert \textbf{V}(i)-\textbf{V}(j)\Vert = |i-j|
\end{equation}
is a deterministic variable, and therefore the correlation sum can be explicitly calculated as
\begin{equation}
\sum_{i<j} \Theta (|i-j|-r)\sim r \Longrightarrow C_1(r)\sim r\;.
\end{equation}
Now, for an arbitrary embedding dimension $m$, the embedded vectors are of the form:
\begin{equation}
\textbf{V}(i)=(i,i+1,i+2,i+3,\dots,i+m-1)\;,
\end{equation}
and according to the $L_{\infty}$ norm we obtain: 
\begin{eqnarray}
\Vert \textbf{V}(i)-\textbf{V}(j)\Vert &=& \max \{ |(i+l)-(j+l)|;l=0,...,m-1 \}\nonumber
\\
&=&|i-j|\;.
\end{eqnarray}
Thus, the arbitrary $m$ case reduces to the case $m=1$, so that $\beta_m =1 \ \forall m$, showing, under the ballistic assumption, a correlation dimension $\beta=1$. $\square$

\subsubsection{Random walker}

\begin{figure*}
\centering
\includegraphics[width=0.8\columnwidth]{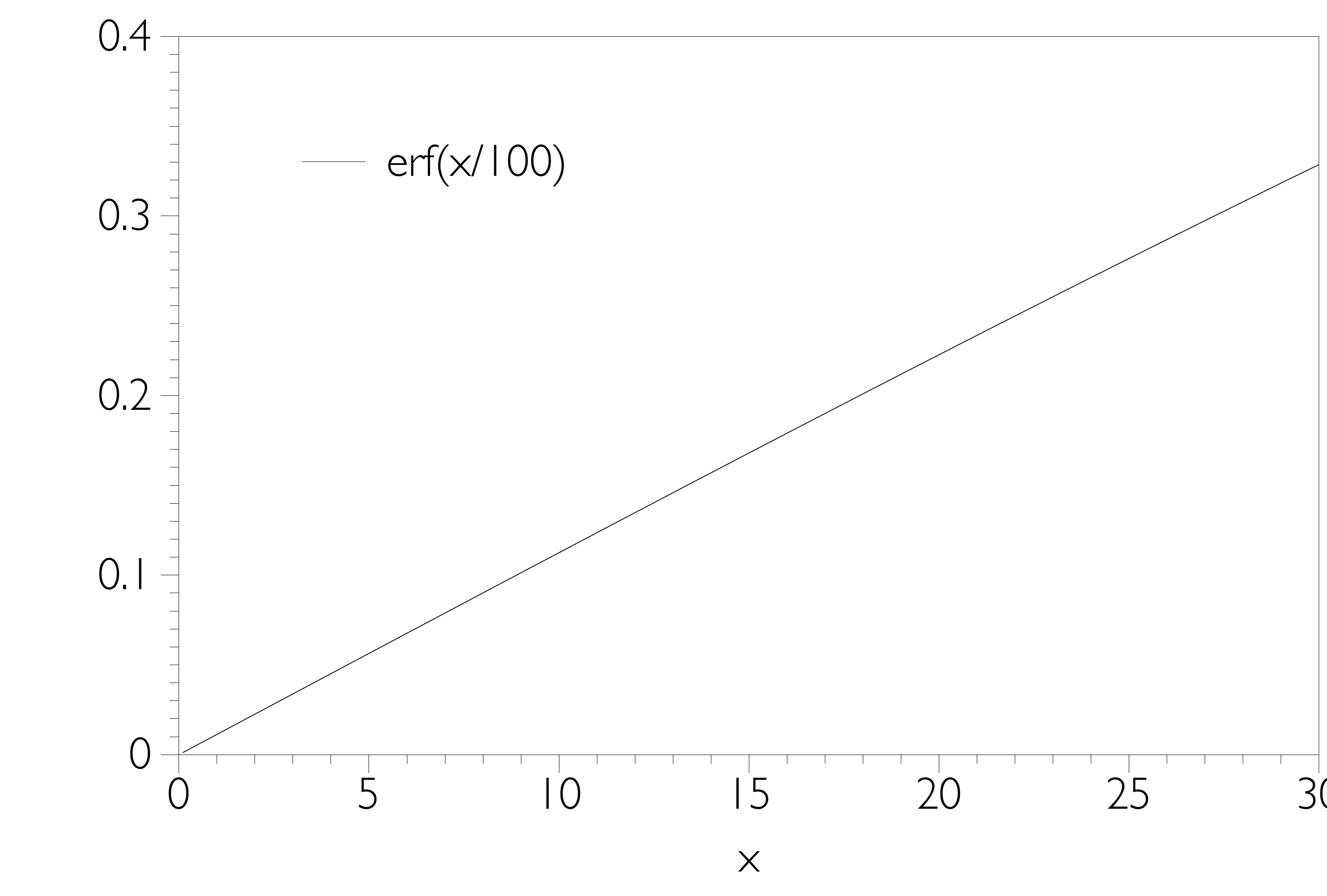}
\includegraphics[width=0.8\columnwidth]{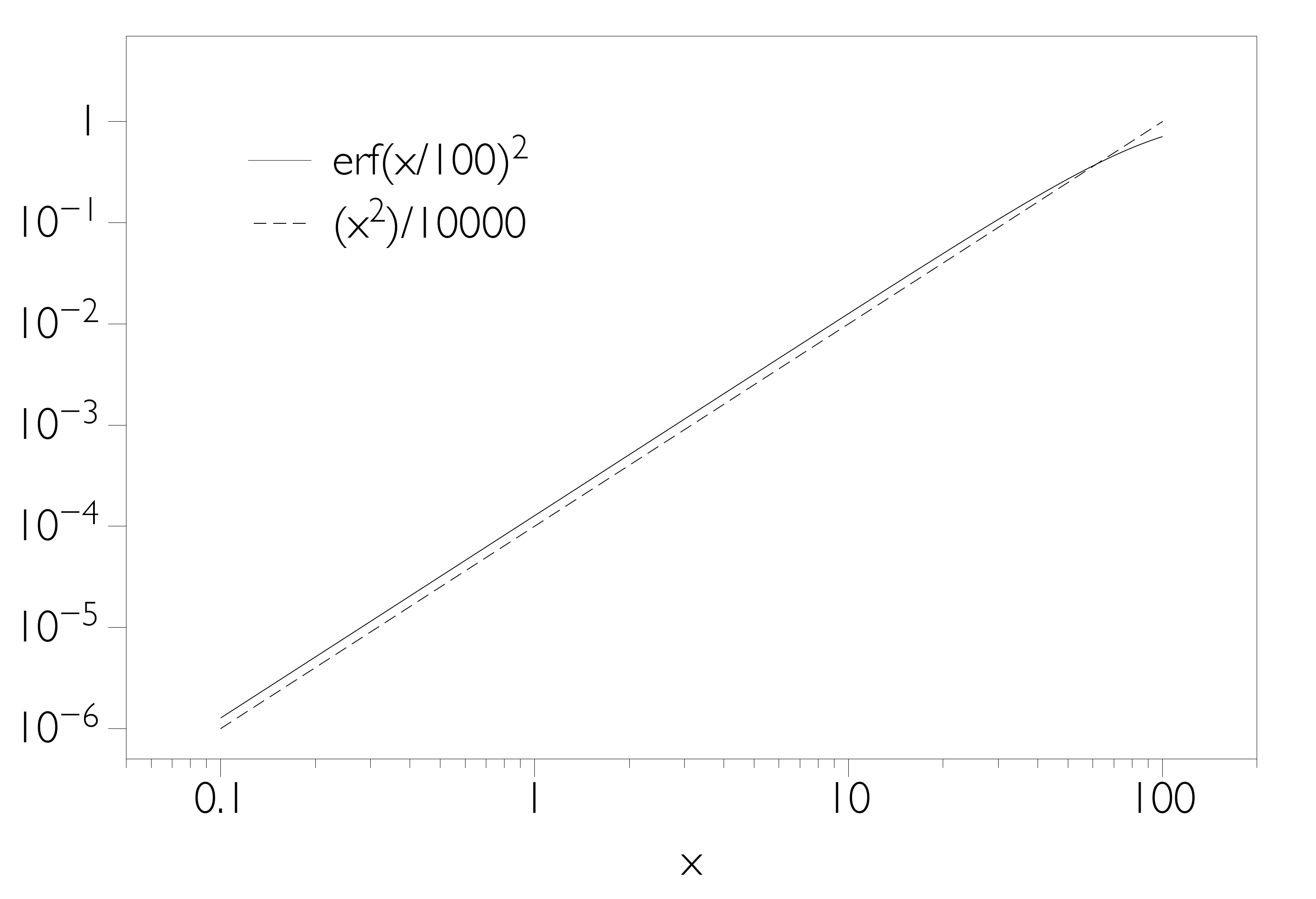}
\caption{ (Left panel) Error function $\text{erf}(x/\sqrt{n})$ for $n=10^4$. For small values of $x$, the function scales linearly with $x$. (Right panel) Log-log plot of $\text{erf}(x/\sqrt{n})^2$ and $x^2/n$ for $n=10^4$. For small values of $x$, both shapes coincide.}
\label{fig_label1}
\end{figure*}

Now we relax the ballistic approximation shown above and present address the correlation dimension derived from the motion of a random walker. First, we label again without loss of generality the nodes of the 1D lattice by consecutive integers, and start by considering
 an embedding dimension $m=1$. In this case the random walker performs a simple walk in $\mathbb{Z}$, and 
\begin{equation}
\Vert \textbf{V}(i)-\textbf{V}(j)\Vert = \xi= |x(i)-x(j)|\;.
\end{equation}
To analyze how $\xi$ is distributed it is easy to notice that the distance between $x(i)-x(j)$ is generated through the sum of $j-i$ random variables, each of which is extracted from $\{-1,+1\}$, which tends to a normal distribution with zero mean and variance $|j-i|$ by virtue of the central limit theorem. Therefore, $\xi$ is the absolute value of the sum of $j-i$ random variables, whose distribution tends to a folded normal distribution with zero mean and variance $|j-i|$. Therefore, after dropping irrelevant constants we obtain:
\begin{equation}
\rho(x)\sim e^{-x^2/(j-i)}\;,
\end{equation}
so that 
\begin{eqnarray}
\sum_{j-i=1}^n\Theta(\xi-r)&=&\sum_{k=1}^n P(\xi_k<r)=\sum_{k=1}^n \int_0^ r \exp(-x^2/k)dx \nonumber \\ 
&=&\sum_{k=1}^n \text{erf}\bigg(\frac{r}{\sqrt{k}}\bigg),
\end{eqnarray}
where $\text{erf}(x)$ is the error function that fulfills:
\begin{equation}
\text{erf}(r/\sqrt{n})\sim r/\sqrt{n}+\mathcal{O}(r^3)
\end{equation}
whose first order is $r$ for $r \leq 10^{-3} n$ (see the left panel of Fig. \ref{fig_label1}),
and therefore:
\begin{equation}
C_1(r)=\sum_{k=1}^n \text{erf}\bigg(\frac{r}{\sqrt{k}}\bigg)\approx \sum_{k=1}^n \frac{r}{\sqrt{k}}\sim r + \text{h.o.t.}\;,
\end{equation}
{\em i.e.}, up to first order $\beta_1=1$ for sufficiently large $n$ and sufficiently small $r$.\\

As a second step, consider an embedding dimension $m=2$. In this situation,
$$\Vert \textbf{V}(i)-\textbf{V}(j)\Vert = \xi= \max \{|x(i)-x(j)|,|x(i)-x(j)\pm 2|\}.$$
Now, the important point is that these three random variables are completely correlated: they are not independent realizations but, on the contrary, all three depend on a single realization of the duple $\{x(i),x(j)\}$. Therefore, we do not need to apply order statistics in this case: $\xi$ is again folded-normally distributed. The argument then proceeds as for $m=1$ such that $C_2(r)\sim r + \text{h.o.t.}$. 

A similar argument holds for a general embedding dimension,$m$, and therefore we conclude that for a 1D lattice, an unbiased random walker generates a correlation sum which, in an embedding dimension $m$ reads:
\begin{equation}
C_m(r)\sim r + \text{h.o.t.}\;,
\end{equation}
that is to say, up to first order the predicted correlation dimension of the 1D lattice is again $\beta=1$. $\square$
%\subsection{That's a subsection}
%\subsubsection{That's a sub-subsection}

\subsection{Lattice $2D$}

We now consider a random walker in a 2D lattice. This is a regular network where all nodes have degree $k_i=4$ that tiles $\mathbb{Z}^2$. In what follows we prove that, up to first order, the correlation dimension of this network is $\beta=2$.\\

In this case each node of this lattice is labelled by a two dimensional vector $(x,y)$, where $x,y \in \mathbb{Z}$. Accordingly, a random walker generates a trajectory of the form
$$\bigg \{ \left(
\begin{array}{c}
x(i)\\
y(i)\\
\end{array}
\right), \left(
\begin{array}{c}
x(i+1)\\
y(i+1)\\
\end{array}
\right),\left(
\begin{array}{c}
x(i+2)\\
y(y+2)\\
\end{array}
\right),\dots,\left(
\begin{array}{c}
x(n)\\
y(n)\\
\end{array}
\right)\bigg \}$$
where the initial $x(i)$ and $y(i)$ are uncorrelated random variables extracted from a uniform discrete distribution $U(1,n)$ and the trajectory is the result of the Markov process defined as:
\begin{equation}
x(i+1)=\bigg \{ 
\begin{array}{c}
x(i)+1,\;{\mbox{with probability}}\;\;1/2\\
x(i)-1, \;{\mbox{with probability}}\;\;1/2
\end{array}
\end{equation}
and
\begin{equation}    
y(i+1)=\bigg \{ 
\begin{array}{c}
y(i)+1,\;{\mbox{with probability}}\;\;1/2\\
y(i)-1,\;{\mbox{with probability}}\;\;1/2\\
\end{array}
\end{equation}
Let us begin analyzing the case of embedding dimension $m=1$. In this case: 
\begin{eqnarray}
\Vert \textbf{V}(i)-\textbf{V}(j)\Vert &=& \xi= \max \{|x(i)-x(j)|,|y(i)-y(j)|\}\nonumber
\\
&=&\max \{\eta_1,\eta_2\},
\end{eqnarray}
where $\eta_1$ and $\eta_2$ are random variables with a probability distribution $f(x)$ which reduces to the case of a 1D lattice, {\em i.e.}, by dropping irrelevant terms:
\begin{eqnarray}
f(x)&\sim& \exp\bigg(\frac{-x^2}{|j-i|}\bigg)\;,\\  
F(x)&\sim&\text{erf}\bigg(\frac{x}{|j-i|^{1/2}}\bigg)\;.
\end{eqnarray}
Therefore, according to order statistics, we find that: 
\begin{equation}
\xi \sim f(x)F(x)= \exp\bigg(\frac{-x^2}{|j-i|}\bigg)\text{erf}\bigg(\frac{x}{|j-i|^{1/2}}\bigg)\;,
\end{equation}
and finally the correlation sum reads:
\begin{eqnarray}
C_1(r)&\sim& P(\xi <r)=\int_0^r \exp \bigg(\frac{-x^2}{|j-i|}\bigg)\text{erf}\bigg(\frac{x}{|j-i|^{1/2}}\bigg) dx \nonumber 
\\
&=& \text{erf}\bigg(\frac{r}{|j-i|^{1/2}}\bigg)^2 \sim r^2 + \text{h.o.t.}\;,
\end{eqnarray}
{\em i.e.}, up to a first order expansion in $r$, the correlation sum for $m=1$, $C_1(r)$, scales quadratically (see the right panel of Fig. \ref{fig_label1} for a numerical check).

Finally, in the general case $m>1$, one can trivially follow an argument similar to the one used for a random walker in a 1D lattice, finding that:
\begin{equation}
C_m(r)\sim r^2 + \text{h.o.t.}, \ \forall m\;,
\end{equation}
{\em i.e.}, the exponent $\beta_m$ saturates to the correlation dimension $\beta=2$. $\square$

\subsection{Lattice $dD$}
To round off, now we prove that in the general case of integer lattices (for a general value $d$), the correlation dimension of the lattice coincides, up to first order, with the Haussdorff dimension of the coarsely equivalent Euclidean space $\beta=d$.\\

First, the trajectory generated by the walker in a $d$ dimensional lattice, where each node is labelled by a d dimensional vector $(x_1,x_2,\dots,x_d), \ x_i \in \mathbb{Z} \ \forall i=1,2,\dots,d$, is: 
\begin{equation}
\left(
\begin{array}{c}
x_1(i)\\
x_2(i)\\
x_3(i)\\
\dots \\
x_d(i)\\
\end{array}\right );
\left(
\begin{array}{c}
x_1(i+1)\\
x_2(i+1)\\
x_3(i+1)\\
\dots \\
x_d(i+1)\\
\end{array}\right );\dots;
\left(
\begin{array}{c}
x_1(n)\\
x_2(n)\\
x_3(n)\\
\dots \\
x_d(n)\\
\end{array}\right )
\end{equation}
and therefore, for a one dimensional embedding ($m=1$) we have
\begin{equation}
\Vert \textbf{V}(i)-\textbf{V}(j)\Vert = \xi=\max \{\eta_1,\eta_2\,\dots,\eta_d\},
\end{equation}
where $\eta_l=|x_l(i)-x_l(j)|$ are random variables with a probability distribution $f(x)$. Finding the probability density of $\xi$ is again an extreme value problem, where order statistics predicts:
\begin{equation}
\xi\sim f(x)F(x)^{d-1}\sim \exp\bigg(\frac{-x^2}{|j-i|}\bigg)\text{erf}\bigg(\frac{x}{|j-i|^{1/2}}\bigg)^{d-1}\;.\end{equation}
Therefore, the correlation sum for $m=1$ reads:
\begin{eqnarray}
C_1(r)&\sim& P(\xi <r)\nonumber\\
&=&\int_0^r \exp \bigg(\frac{-x^2}{|j-i|}\bigg)\text{erf}\bigg(\frac{x}{|j-i|^{1/2}}\bigg)^{d-1} dx \nonumber
\\
&=& \text{erf}\bigg(\frac{x}{|j-i|^{1/2}}\bigg)^d \sim r^d + \text{h.o.t.}\;,
\end{eqnarray}
up to a first order expansion in $r$.
In the general case $m>1$, an argument similar to the one used for a random walker in a 1D lattice holds, thus finding that indeed
\begin{equation}
C_m(r)\sim r^d + \text{h.o.t.}, \; \forall m\;,
\end{equation}
{\em i.e.}, the correlation sum scales with $D$ and thus the correlation dimension of a dD lattice is $\beta=d$. $\square$

\section{Conclusion}

Recently, the notion of fractal dimensionality has been investigated numerically within networks \cite{boxcovering, boxcovering2, boxcovering3, PRL}. The techniques used  have borrowed concepts from measure theory and dynamical systems such as the capacity and correlation dimension respectively. To this aim the corresponding techniques, such as the classical box-counting algorithm and the Grassberger-Procaccia method, have been generalized to the network realm. 

In this manuscript we have focused on the latter of these techniques to show that the correlation dimension of some synthetic networks, as defined in \cite{PRL} and in equations \ref{CS} and \ref{beta}, coincides with the Haussdorff dimension of their coarsely equivalent Euclidean spaces \cite{coarse}. Note that a network and an Euclidean space are very different objects in the small-scale (their topology is entirely different) but they resemble each other in the large-scale. Therefore, our results although desired and expected, are nontrivial. 

In addition, the analytical calculations shown in this manuscript illustrate the validity of the numerical results shown in \cite{review} in more sophisticated synthetic and real-world network. However, finding similar analytical evidences in the case of empirical networks is quite a difficult task. A slightly easier problem which is left for future work is to address the correlation dimension of spatially embedded complex network ensembles with robust statistical properties, {\em i.e.}, the so-called annealed graphs \cite{annealed,annealed1,annealed2,annealed3}.\\

\noindent {\bf Acknowledgments.} The authors would like to thank Pablo Iglesias for inspiring discussions. J.G.G. is supported by MICINN through the Ramon y Cajal program.

\bibliography{apssamp}% Produces the bibliography via BibTeX.

\end{document}